\documentstyle[12pt]{article}
\newcommand{\draft}{
        \renewcommand{\baselinestretch}{1.0}%
        \small\normalsize%
}

\draft
\begin{document}
\title{\bf Dependences of the X-ray luminosity and pulsar wind nebula on 
different parameters of pulsars and the evolutionary effects}
\author{Oktay H. Guseinov$\sp{1,2}$
\thanks{e-mail:huseyin@gursey.gov.tr},
A\c{s}k\i n Ankay$\sp1$
\thanks{e-mail:askin@gursey.gov.tr}, \\
Sevin\c{c} O. Tagieva$\sp3$
\thanks{email:physic@lan.ab.az}, 
M. \"{O}zg\"{u}r Ta\c{s}k\i n$\sp4$
\thanks{email:ozgur@astroa.physics.metu.edu.tr}
\\ \\
{$\sp1$T\"{U}B\.{I}TAK Feza G\"{u}rsey Institute} \\
{81220 \c{C}engelk\"{o}y, \.{I}stanbul, Turkey} \\
{$\sp2$Akdeniz University, Physics Department,} \\ 
{Antalya, Turkey} \\
{$\sp3$Academy of Science, Physics Institute,} \\
{Baku 370143, Azerbaijan Republic} \\
{$\sp4$Middle East Technical University,} \\ 
{Physics Department, 06531 Ankara, Turkey}
}

\date{}
\maketitle
\begin{abstract}
\noindent
Dependences of the X-ray luminosity (L$_x$) of young single pulsars, due 
to ejection of relativistic particles, on electric field intensity, rate 
of rotational energy loss (\.{E}), magnetic field, period, and some other 
parameters of neutron stars are discussed. Influence of the magnetic 
field and effects of some other parameters of neutron stars on the 
L$_x$-\.{E} and the L$_x$-$\tau$ (characteristic time) dependences are 
considered. Evolutionary factors also play an important role in our 
considerations. Only the pulsars with L$_{2-10keV}$$>$10$^{33}$ erg/s 
have pulsar wind nebula around them. The pulsars from which $\gamma$-ray 
radiation has been observed have low X-ray luminosity in general.     
\end{abstract}

\section{Introduction}
X-ray luminosity (L$_x$) of single neutron stars 
strongly depends on the rate of rotational energy loss (\.{E}) $^{1,2,3}$.
\.{E} depends on the spin period (P) and time derivative of the spin 
period (\.{P}) of neutron stars. As there exist many other parameters of 
pulsars determined from P and \.{P} values, L$_x$ must also 
depend on these parameters. This problem has already been analysed  
using values of L$_x$ in 2-10 keV band for different groups of pulsars 
including also the old millisecond pulsars $^2$. We also
prefer to use the L$_x$ values in 2-10 keV band, because the 
absorption is strong in the softer X-ray band and also there must be a 
contribution of the cooling radiation to the soft X-ray part of the 
radiation of pulsar which may complicate the 
problem. Below, using reliable distance values we have analysed in 
detail the dependence of L$_{2-10keV}$ on different pulsar parameters for 
young single pulsars for which the radiation is related to ejection of 
relativistic particles. In this investigation we have also used 
the values of
pulsar radiation in 0.1-2.4 keV band and the radiation of pulsar wind 
nebulae (PWNe) in both bands. Special attention has been paid to the 
evolutionary factors. There exist many pulsars with large values of \.{E} 
and small values of age from which no X-ray radiation has been observed. 
Below, we have also discussed possible reasons of this.     

\section{Dependences of X-ray radiation of pulsars and their wind nebula 
on different parameters of pulsars} 
For the pulsed radio emission to occur and in order to produce radiation 
(excluding cooling radiation) in X-rays and also to form 
PWN, there must be ejection of high-energy particles. 
The production of these kinds of radiation and PWN must depend on  
different parameters of neutron star. Note that 
the magnetic field may not be pure magneto-dipole field and the angle 
between the magnetic field and the spin axis is uncertain. These facts  
also make the solution of the problems more difficult. We want to 
figure out whether X-ray radiation of pulsars and the formation of PWN 
depend on \.{E} or the electric field intensity (E$_{el}$) more strongly 
for the case of pure magneto-dipole radiation. In order to clarify these 
problems, we have constructed X-ray luminosity (2-10 keV) versus \.{E} 
diagram for all radio and single X-ray pulsars with characteristic time 
($\tau$) $<$10$^6$ yr located up to 7 kpc from the Sun including also 2 
pulsars in Magellanic Clouds (Figure 1). The data of these pulsars are 
represented in Table 1. 

Depending on values of the magnetic field and the speed of 
rotation on the surface of neutron star, there arises induced electric 
field of which the intensity is represented as $^4$:
\begin{equation}
E_{el} = \frac{4\pi R}{c} \frac{B_r}{P} \sim (\frac{\dot{P}}{P})^ 
\frac{1}{2}
\end{equation}
where R is the radius of neutron star, B$_r$ the 
value of the real magnetic field strength and c the speed of light. 
In actuality, pulsars are located in plasma and expression (1) does not 
give the exact value of E$_{el}$.
As the P-\.{P} diagram which we use is always represented in             
logarithmic scale, the lines of constant E$_{el}$
pass parallel to the lines of constant characteristic time
\begin{equation}
\tau = \frac{P}{2\dot{P}}
\end{equation}
so that, we do not show the lines of E$_{el}$=constant on the P-\.{P} 
diagram. Calculated values of E$_{el}$ may also contain some mistakes, 
because different pulsars may have different radius, moment of inertia 
(I) and braking index. 

Often, there is PWN around the youngest pulsars which we include in Table 
1. For such cases the X-ray luminosity value include both the X-ray 
radiation of the pulsar and of the PWN in most of the cases, as it is 
difficult to distinguish the L$_x$(PWN) part of the luminosity. On the 
other hand, since the contribution of the PWN to the X-ray luminosity can 
be at most comparable with the X-ray luminosity of the pulsar in general 
$^3$, the change in the positions of the pulsars in the figures will be  
small in such cases that the dependences found from the best fits do not 
change considerably. It must also be noted that the uncertainties in the 
distance values and in the measured 2-10 keV fluxes can not have 
significant influences on these dependences. The pulsars with 
$\tau$$<$10$^6$ yr located up to 7 kpc for which the X-ray luminosity 
only in the 0.1-2.4 keV band is known are also displayed in Table 1. We 
have adopted reliable values of distance for the pulsars $^5$ and X-ray 
luminosity data mainly have been taken from Possenti et al. $^2$ and 
Becker \& Aschenbach $^3$. 

The dependence between X-ray luminosity (2-10 keV) and \.{E} is displayed 
in Figure 1. The equation of this dependence is:
\begin{equation}
L_{2-10keV} = 10^{-23.40\pm 4.44} \dot{E} ^{1.56\pm 0.12}. 
\end{equation}
In Figure 2, X-ray luminosity (2-10 keV) versus $\tau$ diagram is 
represented for the same pulsar sample shown in Figure 1. The equation for 
the relation between X-ray luminosity (2-10 keV) and $\tau$ is:
\begin{equation}
L_{2-10keV} = 10^{41.79\pm 1.04} \tau ^{-1.98\pm 0.23}.
\end{equation}
As seen from Figures 1 and 2, the deviations 
of the data with respect to the best fits shown in the figures are larger 
in the L$_x$ (2-10 keV) versus $\tau$ diagram compared to the L$_x$ (2-10 
keV) versus \.{E} diagram, but there is no significant difference. 

Some pulsars in Table 1 have $\gamma$-ray radiation. As seen from Figure 
1, these pulsars are in general located below the best fit line. On the 
other hand, all the pulsars with large effective magnetic field (B) values 
(log B $>$ 12.7) are located above the line (note that when there exist 
some additional mechanisms other than the magneto-dipole mechanism, 
B$>$B$_r$). We have not designated these pulsars in Figure 
2 but notice that $\gamma$-ray pulsars in this figure are located in both 
parts with respect to the best fit line, whereas, all the pulsars with 
large values of B have positions below the line. This is also the result 
of different dependences of E$_{el}$ (see expression (1)) and rate of 
rotational energy loss
\begin{equation}
\dot{E}=\frac{4\pi ^2 I \dot{P}}{P^3},
\end{equation}
on the value of P (note that in this approximation we neglect the 
dependences of the observational and the calculated parameters of pulsars 
on the radius and mass of neutron stars). 

Which parameters of neutron stars do the deviations in Figures 1 and 2 
mainly depend on? In order to understand this, we have examined pulsars 
in narrow \.{E} intervals with large differences in their L$_x$ values 
which lead to large deviations. The largest deviation is for the 
interval log \.{E}=36.78-36.9 where pulsars J1846-0258, J1811-1925, and 
Vela are present. These are not 
ordinary pulsars: pulsars J1846-0258 and J1811-1925 are strong X-ray 
pulsars which have not been detected in other bands, whereas, Vela pulsar, 
which has been known as a radio, optical and gamma-ray pulsar until 
recently, has been identified also in X-rays but with low luminosity $^6$. 
The L$_x$ (2-10 keV) value of Vela pulsar (Table 1) is more 
than one order of magnitude larger than the value given by Possenti 
et al. $^2$. 
This is because we have also included the X-ray luminosity of the 
PWN around the Vela pulsar which is about 10 times greater than the X-ray 
luminosity of the pulsar $^7$ and we have adopted a 
more reliable distance value $^5$ so that the luminosity 
value also increased about 3 times. Despite these facts, the position of 
Vela pulsar in Figure 1 is still well below the line and the difference 
between the luminosity values of Vela pulsar and pulsar J1846-0258 is 
still very large. Although, the period value of pulsar J1846-0258 is 4-5 
times larger than the period values of Vela pulsar and pulsar 1811-1925,
since the magnetic field 
\begin{equation}
B=\sqrt{\frac{3c^3IP\dot{P}}{8\pi ^2 R^6}}
\end{equation}
and \.{E} values of pulsar J1846-0258 are larger 
than the values 
of the other two pulsars (even though the $\tau$ value is small),  
its luminosity (2-10 keV) value is larger (see Table 1). 

The second largest deviation among large \.{E} 
values in Figure 1 is in the log \.{E}=37.21-37.42 interval where pulsars 
J1513-5908, J0205+6449, J1617-5055, and J2229+6114 are located in. Pulsar 
J1513-5908, the youngest one among these pulsars, has the largest P, B and 
L$_x$ values. Among the other pulsars in this group, the largest \.{E} 
value belongs to pulsar J0205+6449 in this 
interval and this pulsar has a smaller value of $\tau$, a larger value of 
B, and a period value in between the values of pulsars J1617-5055 and 
J2229+6114. In spite of this, the luminosity of this pulsar is less than 
the luminosity of pulsar J1617-5055. This is strange, because all 
four parameters (P, $\tau$, B, \.{E}) of pulsar J0205+6449 suggest a 
larger luminosity value. Furthermore, there is no supernova shell or PWN 
around pulsar J1617-5055. This contradiction is not because of the 
adopted distance values. Is this 
contradiction related to the uncertainties in the observational data?  
Although, the luminosity values of these 2 pulsars (J0205+6449 and 
J1617-5055) are comparable within error limits $^2$, one 
still can not explain why pulsar J0205+6449 does not have a larger L$_x$ 
value. If the observational data are not significantly different than the 
actual values, then either there may be a considerable difference in the 
mass values of these 2 pulsars or the magnetic field may not be pure 
dipole.  

Among the youngest pulsars, the largest deviations are in the log 
$\tau$=3.19-3.22 interval. Among the pulsars in this interval, the 
smallest P and the largest \.{E} values belong to pulsar J0540-6919 which 
has also the largest L$_x$ value. In this interval, the smallest L$_x$ 
value 
belongs to pulsar J1119-6127 which has the largest P and B values and the 
smallest \.{E} value. According to Gonzalez \& Safi-Harb $^8$, J1119-6127 
has unabsorbed L$_x$ (0.5-10 keV)=5.5$^{+10}_{-3.3}$$\times$10$^{32}$ 
erg/s at 6 kpc. There is no pulsed X-ray radiation observed from radio 
pulsar J1119-6127 and the position of the X-ray source is not coincident 
with the position of the radio pulsar; the observed point source may 
actually be a PWN $^9$. 

Pulsar J0358+5413 has log $\tau$=5.75 and pulsar 
J0538+2817 has log $\tau$=5.79. The L$_x$ value of pulsar J0358+5413 is 
about 300 times larger than the L$_x$ value of pulsar J0538+2817, but the 
P, B and \.{E} values of these 2 pulsars are roughly the same. Why is 
pulsar J0358+5413 much more luminous than pulsar J0538+2817? The absolute 
error of the L$_x$ (2-10 keV) value of pulsar J0358+5413 is very large 
$^2$ and the L$_x$ (2-10 keV) value is not a directly 
measured value but converted by Possenti et al. $^2$ from the L$_x$ 
value in softer band $^{10}$. So, the high L$_x$ (2-10 keV) 
value of pulsar J0358+5413 can actually be much lower and the large 
difference in the L$_x$ (2-10 keV) values of pulsars J0358+5413 and 
J0538+2817 can actually be much smaller. 

If we consider the corrections to the positions of some of the pulsars 
discussed above taking also into account relations between L$_x$ and 
\.{E}, $\tau$, B and P, and if we change the positions of the pulsars in 
the figures within error limits given in Possenti et al. $^2$ then, 
pulsar J1119-6127 will have a higher L$_x$ value, whereas, pulsars 
J1826-1334, 
J1302-6350 and J0358+5413 will have lower values of L$_x$. In this case, 
the dependence between L$_x$ and $\tau$ given in eqn.(4) will be improved 
with a more negative power of $\tau$. If we apply the same approach for 
the positions of some of the pulsars in Figure 1, we see that the 
deviations do not decrease and the power of \.{E}  
(eqn.(3)) increases negligibly. 
 
Some of the pulsars shown in Figures 1,2 and in Table 1 have experienced 
strong glitches ($\frac{\Delta P}{P}$$>$10$^{-6}$, denoted with 'G' in 
Figure 2). As seen from Figure 2, the deviations of the data do not 
seem to be due to the glitch activity. 

\section{The evolutionary factors}
Since the values of P and \.{P} and the other parameters change 
continuously during the pulsar evolution, the value of L$_{2-10keV}$ must 
also change in connection to this. We want to analyse how the 
L$_{2-10keV}$ luminosity of pulsars changes during the evolution of 
pulsars on the P-\.{P} diagram.

From eqn. (3) we see that
\begin{equation}
L_{2-10keV}^{(\dot E)} \propto (\frac{\dot{P}}{P^3})^{1.56}
\end{equation}
and similarly from eqn. (4)
\begin{equation}
L_{2-10keV}^{(\tau)} \propto (\frac{\dot{P}}{P})^{1.98}
\end{equation}
using the expressions for \.{E} (5) and $\tau$ (2). Now, using 
the expression for B (6), we can write (7) and (8) in the form:
\begin{equation}
L_{2-10keV}^{(\dot E)} \propto (\frac{B^2}{P^4})^{1.56}
\end{equation}
and
\begin{equation}
L_{2-10keV}^{(\tau)} \propto (\frac{B^2}{P^2})^{1.98}.    
\end{equation}
If both dependences (3) and (4) are  
applicable in finding the value of L$_x$, then the ratio of the 
luminosities for each pulsar found from equations (9) and (10)
must roughly be the same throughout the evolution of a pulsar. 
From dependences (10) and (9) we get:
\begin{equation}
\frac{L_{2-10keV}^{(\tau)}}{L_{2-10keV}^{(\dot E)}} \propto B^{0.84} 
P^{2.28}.
\end{equation}
So, the ratio is strongly dependent on P which must be wrong because the 
ratio must not change significantly during the evolution if both 
dependences are reliable. 

X-ray luminosity of the considered pulsars strongly depends on different 
parameters of pulsars and decreases with age down to 
L$_x$$\sim$10$^{29}$-10$^{30}$ erg/s. During the evolution of these 
pulsars, the period value changes up to a factor of 10, \.{P} up to 50 
times, \.{E} up to 10$^4$ times, and $\tau$ up to 500 times. During the 
evolution, value of B for each pulsar may decrease only up to a factor of 
3, but magnitudes of the deviation of B values for different pulsars are 
larger. 

As seen from Figure 5, an increase in the initial magnetic field value of 
a factor of 10 leads to a decrease in the age of X-ray pulsar of a 
factor of 10. As X-ray luminosity of young pulsars strongly depends on 
E$_{el}$ and \.{E}, 
as seen from the L$_x$-$\tau$ diagram (Figure 2) and Figure 5, the X-ray 
luminosity of the pulsars with largest initial B values drop very 
quickly below the threshold during the evolution, whereas, the X-ray 
luminosity of the pulsars with smaller initial values of B last longer 
and drop below the threshold in about 10$^5$ yr. Therefore, it is 
necessary to consider a more homogeneous group of pulsars which have B 
values close to each other. Naturally, the best fit lines for the new 
sample must lead to a different equation for the L$_x$-$\tau$ dependence 
and practically the same equation for the L$_x$-\.{E} dependence.        
In order to see this, we have 
constructed L$_x$ vs. \.{E} and L$_x$ vs. $\tau$ diagrams (Figures 3 and 
4) for the pulsars with B values in the interval 12.2 $\le$ log B $\le$ 
12.7 (Figure 5). Again from the best fits, we get dependences 
between L$_x$ and \.{E}:
\begin{equation}
L_{2-10keV} = 10^{-27.46\pm 6.48} \dot{E} ^{1.66\pm 0.18}
\end{equation} 
and L$_x$ and $\tau$:
\begin{equation}
L_{2-10keV} = 10^{45.98\pm 1.51} \tau ^{-2.99\pm 0.36}.
\end{equation}
From these two dependences it is seen that the dependence given by eqn.(3) 
is comparable to the dependence given by eqn.(12) within error limits, 
but the dependence given by eqn.(4) changed considerably (see eqn.(13)). 

Dividing eqn.(13) by eqn.(12) and again using the expressions (5) and (2) 
we get:
\begin{equation}
\frac{L_{2-10keV}^{(\tau)}}{L_{2-10keV}^{(\dot E)}} \propto B^{2.66}
P^{0.66}.
\end{equation}
As the effective B value decreases gradually during the evolution, the 
large increase in the P value compensates it and the ratio remains 
approximately the same. The dependences (12) and (13) are more reliable 
and reflect the evolutionary effects, because we have used a more 
homogeneous group of pulsars with similar evolutionary tracks in Figures 
3 and 4. The ratio given by dependence (14) shows this fact. 

The reason of the considerable difference between the dependences (4) and 
(13) can easily be seen in Figure 5. As seen from this figure, when the 
value of $\tau$ is small, first the constant $\tau$ line 
crosses the considered B interval (i.e. the interval of 12.2 $\le$ log B 
$\le$ 12.7) where the value of \.{E} is large. Then, it crosses the 
region, which is not included in the B interval under consideration,
where the value of \.{E} is small. On the other hand, if the value of 
$\tau$ is large, then the 
constant $\tau$ line first crosses a region, outside the 
considered B interval, where pulsars with large values of \.{E} are 
located, and then it crosses the B interval under consideration where 
pulsars with small \.{E} values are present. Therefore, for small 
values of $\tau$, the pulsars which have smaller values of L$_{2-10keV}$ 
and which are not located in the considered interval have positions below 
the best fit line in Figure 4. On the other hand, for large 
values of $\tau$, the pulsars which have larger values of L$_{2-10keV}$ 
and which are not located in the considered interval have positions above 
the best fit line in Figure 4. Excluding the pulsars with large deviations 
from the best fit line leads to the considerable difference between the 
dependences (4) and (13). 

In Table 1, we have also included the 15 pulsars with $\tau$$<$10$^5$ yr 
and/or log \.{E}$>$35.60 from which no X-ray radiation has 
been detected. In Figure 5, all the 48 pulsars in Table 1 are 
displayed. Within and around this region of the P-\.{P} diagram, 
there are pulsars with detected X-ray radiation. All of these 
pulsars without detected X-ray radiation are also located at d$\le$7 kpc 
from the Sun. One of these pulsars is connected to a supernova remnant 
(SNR). 

As seen from Figure 5 and Table 1, there are some pulsars without observed 
X-ray radiation which have smaller $\tau$ values and larger \.{E} values  
compared to some pulsars with observed X-ray radiation. What is the reason 
for not detecting X-ray radiation from such pulsars? As mentioned in 
the previous section, the X-ray luminosity must depend not only on P and 
\.{P} but 
also on some other parameters. The deviations of the data from the best 
fits in Figures 1-4 also show this fact. It is not easy to examine 
these "other parameters" which can not be calculated from values of P 
and \.{P} and get realiable results. But it is possible to 
clarify why there exist those pulsars with smaller $\tau$ values and 
larger \.{E} values from which no X-ray radiation has been detected by 
examining the selection effects. 

It is clear that it is more difficult to 
observe the X-ray radiation of a pulsar if it is located at a large 
distance in the Galactic plane and in the Galactic centre direction. 
All the 8 pulsars with $\tau$$>$10$^5$ yr from which X-ray radiation 
has been observed as shown in Figure 5 are located at distances d$<$2.5 
kpc from the Sun (Figure 6). Directions of these and all the other 
pulsars in Table 1 are shown in b vs. l diagram (Figure 7). 
 
In Figure 6, \.{E}--d diagram for all the 48 pulsars in 
Table 1 is represented. The pulsars without observed X-ray radiation are 
all located beyond 2 kpc from the Sun. It is normal to detect 
X-ray radiation from pulsars with large \.{E} values even at 
large distances (see Figure 6), but there is a pulsar, namely J0631+1036, 
with detected X-ray radiation in the 2-10 keV band which is located at 
d=6.6 kpc and has a smaller \.{E} value (log \.{E}=35.24). The reason for 
this is that this pulsar is located in the Galactic anti-centre direction 
(see Figure 7). The other 8 pulsars with detected X-ray radiation 
and with distances between 6-7 kpc are located in the Galactic central 
directions. 
Six out of the 8 far away pulsars in the central directions are connected 
to SNRs that these are well examined pulsars. So, it must be considered 
normal to detect the X-ray radiation of these pulsars. 
On the other hand, the remaining 2 pulsars, namely J1105-6107 and 
J1617-5055, are not connected to SNRs. Pulsar J1617-5055 has 
very high \.{E} and large L$_x$ values (see Table 1), and pulsar 
J1105-6107 is located in between the Sagittarius and the Scutum arms 
so that it is normal to observe the 
X-ray emission of these 2 pulsars. All the pulsars without detected X-ray 
radiation are located in the central directions and in the directions 
of Vela OB-associations (see Figure 7). Moreover, the pulsars with 
d$\le$4.1 kpc which have not been detected in X-rays  
(J0940-5428, J1809-1917, J1718-3825, J1531-5610, J1509-5850, J1913+1011) 
have all been found in recent pulsar surveys $^{11,12,13}$.
In the near future, it is possible that all of these pulsars will be 
detected in X-rays. 

\section{Discussion and Conclusions}
For single pulsars, the X-ray radiation due to the relativistic particles, 
which considerably exceeds the cooling radiation, is important and worth 
examining. Power and spectrum of this radiation depend on the number and 
the energy spectra of the relativistic electrons in the magnetospheres of 
pulsars. They also depend on the number density of charged particles and 
the magnetic field in the magnetosphere. The X-ray radiation of PWNe also 
depends on these 
quantities. X-ray radiation has also been observed from old single 
millisecond pulsars which is absolutely not cooling radiation but a result 
of Coulomb interaction between the accelerated particles and the charged 
particles in the magnetosphere. Since the parameters of millisecond 
pulsars practically do not vary in time and as they are not active as 
the young pulsars are, we do not consider them in this work.    

From the investigations of single pulsars which radiate X-rays as a result 
of ejection of relativistic particles, we have concluded as follows: \\
1) Luminosity in 2-10 keV band (often the total luminosity including the 
luminosity of the PWN) strongly depends on electric field intensity 
E$_{el}$ (or $\tau$), rate of rotational energy loss \.{E}, magnetic field 
B, period, and in smaller degree on some other parameters of neutron 
stars. \\
2) The dependence of L$_x$ on \.{E} is practically the same for 
pulsars with very different values of B and with B=10$^{12}$-10$^{13}$ G. 
This is not true for the L$_x$-$\tau$ dependence. \\
3) Pulsars with different initial values of B and with $\tau$ about 
10$^3$ yr begin to evolve from different points of the L$_x$-\.{E} 
dependence and then continue to evolve with similar trajectories until 
L$_x$ drops to $\sim$10$^{30}$ erg/s, but the same pulsars begin and end 
their evolution in different parts of the L$_x$-$\tau$ dependence and 
the evolutionary tracks are not parallel to the L$_x$-$\tau$ dependence 
for all the pulsars under consideration. \\
4) The pulsars, which also radiate gamma-rays, with the same values of 
\.{E} as the other pulsars without gamma-ray radiation have several 
times smaller X-ray luminosities. \\
5) The pulsars which have PWNe practically always have X-ray luminosities 
greater than 10$^{33}$ erg/s. \\ 
6) Practically, values of $\tau$ and \.{E} determine the X-ray 
luminosity. The increase in $\tau$ and the decrease in \.{E} 
simultaneously lead to a rapid decrease in L$_x$. Value of L$_x$ decreases 
from $\sim$10$^{37}$ erg/s for \.{E}$\sim$3$\times$10$^{38}$ erg/s and 
$\tau$$\sim$10$^3$ yr to $\sim$10$^{30}$ erg/s for 
\.{E}$\sim$3$\times$10$^{34}$ erg/s and $\tau$$\sim$3$\times$10$^5$ yr. 

If we take into account these and also the strength and spectra of the 
X-ray radiation of millisecond pulsars $^{2,3}$, we can have some 
preliminary conclusions about the acceleration of particles and their 
radiation. As the characteristic time of millisecond pulsars is about 4-5 
orders of magnitude smaller compared to the young pulsars, their E$_{el}$ 
values must be 200-250 times smaller than the E$_{el}$ values of young 
pulsars. On the other hand, the value of E$_{el}$ does not have a strong 
influence on the X-ray radiation of pulsars. The X-ray spectra of young 
pulsars are, on average, steeper compared to the spectra of millisecond 
pulsars. If we also consider the strong dependence of the X-ray radiation 
of all the pulsars on the value of \.{E}, we can conclude as follows: \\
7) Acceleration of particles mainly takes place in the field of 
magnetodipole radiation wave. E$_{el}$ has a role of only triggering this 
process. \\
8) The high values of X-ray luminosity of young pulsars under the same 
\.{E} values and the steeper spectra of such pulsars are related to the 
large amount of charged particles in their magnetospheres as compared to 
millisecond pulsars.  

It is necessary to reexamine the X-ray flux values of J0205+6449, 
J0358+5413, and J0538+2817 in 2-10 keV band to understand the dependences 
of the X-ray luminosity on \.{E}, E$_{el}$, B and P values and the 
possible dependence on the energy spectra of the ejected particles.  

Also, in order to understand the dependence of the X-ray radiation on 
different parameters of pulsars better, it is necessary to observe pulsars 
J0940-5428, J1809-1917, J1718-3825, J1531-5610, J1509-5850, J1913+1011 
preferably in the hard X-ray band.

\clearpage
\begin{figure}[t]
\vspace{3cm}
\includegraphics{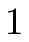}
\end{figure}

\clearpage
{\bf Figure Captions} \\
{\bf Figure 1:} Log L$_x$ (2-10 keV) versus log \.{E} diagram of all 30 
pulsars with $\tau$ $<$ 10$^6$ yr which have observed X-ray radiation. 28 
of these pulsars are located up to 7 kpc 
from the Sun and 2 of them are in Magellanic Clouds. 'Plus' signs denote 
the 2 pulsars in Magellanic 
Clouds. 'Cross' signs show the positions of pulsars with 10$^5$ $<$ 
$\tau$ $<$ 10$^6$ yr and 'star' signs show the positions of pulsars with 
$\tau$ $<$ 10$^5$ yr. 'Square' signs denote single X-ray pulsars.  
Seven of these pulsars have log B $>$ 12.7 (denoted with 'B') and from 8 
of them $\gamma$-rays have been observed (denoted with 'G'). \\
{\bf Figure 2:} Log L$_x$ (2-10 keV) versus log $\tau$ diagram of all 30
pulsars with $\tau$ $<$ 10$^6$ yr which have observed X-ray radiation. 28 
of these pulsars are located up to 7 kpc
from the Sun and 2 of them are in Magellanic clouds. 'Plus' signs denote 
the 2 pulsars in 
Magellanic Clouds. 'Cross' signs show the positions of pulsars with 
10$^5$ $<$ $\tau$ $<$ 10$^6$ yr and 'star' signs show the positions of 
pulsars with $\tau$ $<$ 10$^5$ yr. 'Square' signs denote single X-ray 
pulsars. Seven of these pulsars have experienced strong ($\Delta$P/P $>$ 
10$^{-6}$) glitches (denoted with 'G'). \\
{\bf Figure 3:} Log L$_x$ (2-10 keV) versus log \.{E} diagram for 15 of 
the 
30 pulsars shown in Figures 1 and 2; these 15 pulsars have 12.2 $\le$ log 
B $\le$ 12.7. 'Plus' sign denotes the pulsar in Magellanic Cloud. 'Cross' 
signs show the positions of pulsars with 10$^5$ $<$   
$\tau$ $<$ 10$^6$ yr and 'star' signs show the positions of pulsars with
$\tau$ $<$ 10$^5$ yr. 'Square' sign denotes the single X-ray pulsar. \\
{\bf Figure 4:} Log L$_x$ (2-10 keV) versus log $\tau$ diagram for 15 of 
the 30 pulsars shown in Figures 1 and 2; these 15 pulsars have 12.2 $\le$ 
log B $\le$ 12.7. 'Plus' sign denotes the pulsar in Magellanic Cloud. 
'Cross' signs show the positions of pulsars with 10$^5$ $<$
$\tau$ $<$ 10$^6$ yr and 'star' signs show the positions of pulsars with
$\tau$ $<$ 10$^5$ yr. 'Square' sign denotes the single X-ray pulsar.
Black squares show the positions of the pulsars which are 
present in Figure 2 but not included in the fit of Figure 4. \\
{\bf Figure 5:} Period versus period derivative diagram for all 48 pulsars 
represented in Table 1 with $\tau$ $<$ 10$^6$ yr and distance $\le$ 7 kpc 
including the 2 pulsars in Magellanic Clouds. 'Circles' represent the 33 
pulsars from which X-ray radiation have been detected in 2-10 keV and/or 
0.1-2.4 keV bands. 'Plus' signs denote the 15 pulsars with log \.{E} 
$>$ 35.55 from which no X-ray radiation has been observed. \\ 
{\bf Figure 6:} Log \.{E} versus distance diagram for all 48 pulsars
represented in Table 1 with $\tau$ $<$ 10$^6$ yr and distance $\le$ 7 kpc
including the 2 pulsars in Magellanic Clouds. 'Circles' represent the 33  
pulsars from which X-ray radiation have been detected in 2-10 keV and/or
0.1-2.4 keV bands. 'Plus' signs denote the 15 pulsars with log \.{E}
$>$ 35.55 from which no X-ray radiation has been observed. 'Cross' signs 
represent the 8 pulsars with 10$^5$$<$$\tau$$<$10$^6$ yr.\\
{\bf Figure 7:} Galactic longitude (l) versus Galactic latitude (b) 
diagram for all 48 pulsars represented in Table 1 with $\tau$ $<$ 10$^6$ 
yr and distance $\le$ 7 kpc including the 2 pulsars in Magellanic Clouds. 
'Circles' represent the 33 pulsars from which X-ray radiation have been 
detected in 2-10 keV and/or 0.1-2.4 keV bands. 'Plus' signs denote the 15 
pulsars with log \.{E} $>$ 35.55 from which no X-ray radiation has been 
observed. 'Cross' signs represent the 8 pulsars with 
10$^5$$<$$\tau$$<$10$^6$ yr.

\clearpage  
\begin{figure}[t]
\vspace{3cm}
\includegraphics{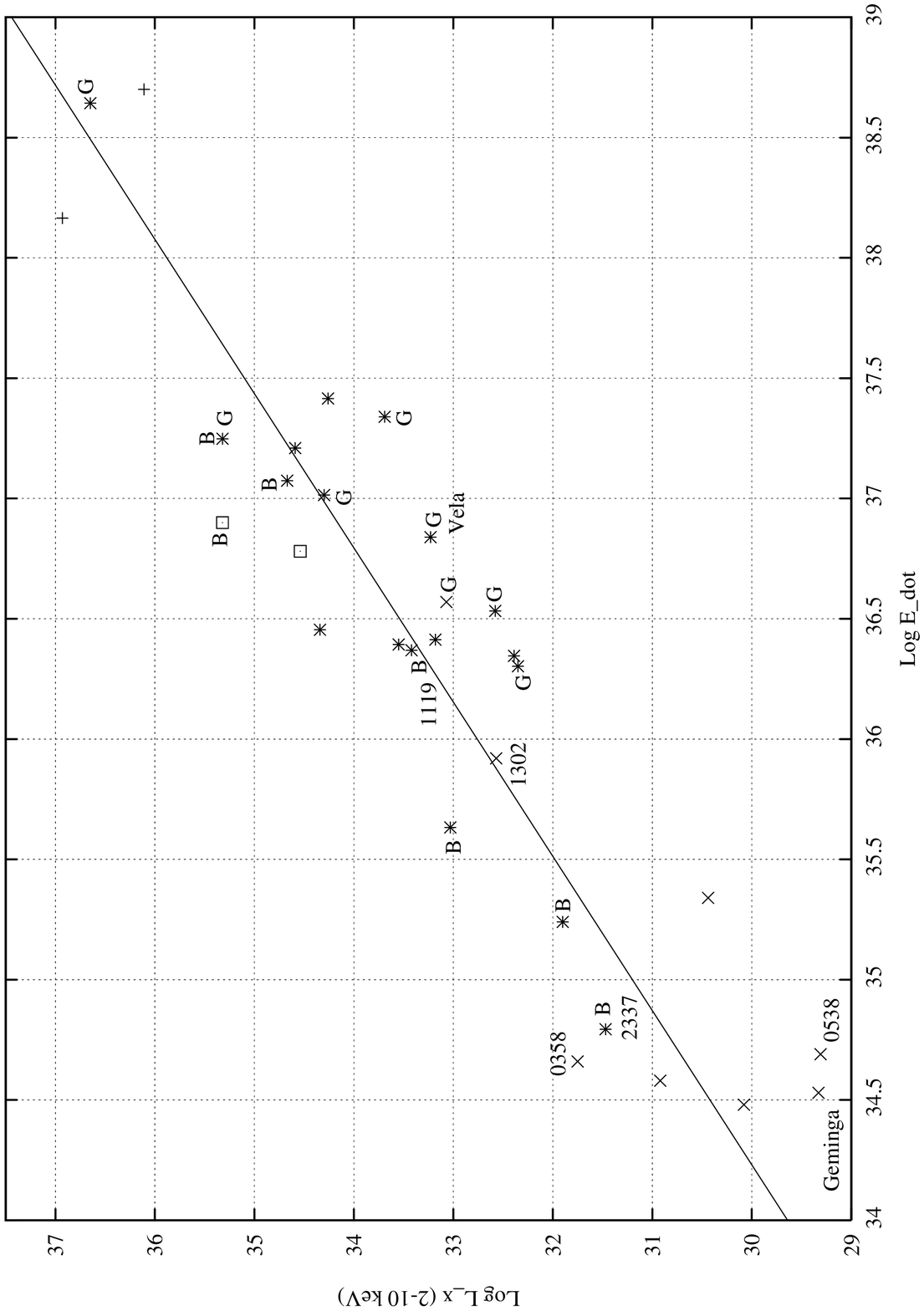}
\end{figure}

\clearpage  
\begin{figure}[t]
\vspace{3cm}
\includegraphics{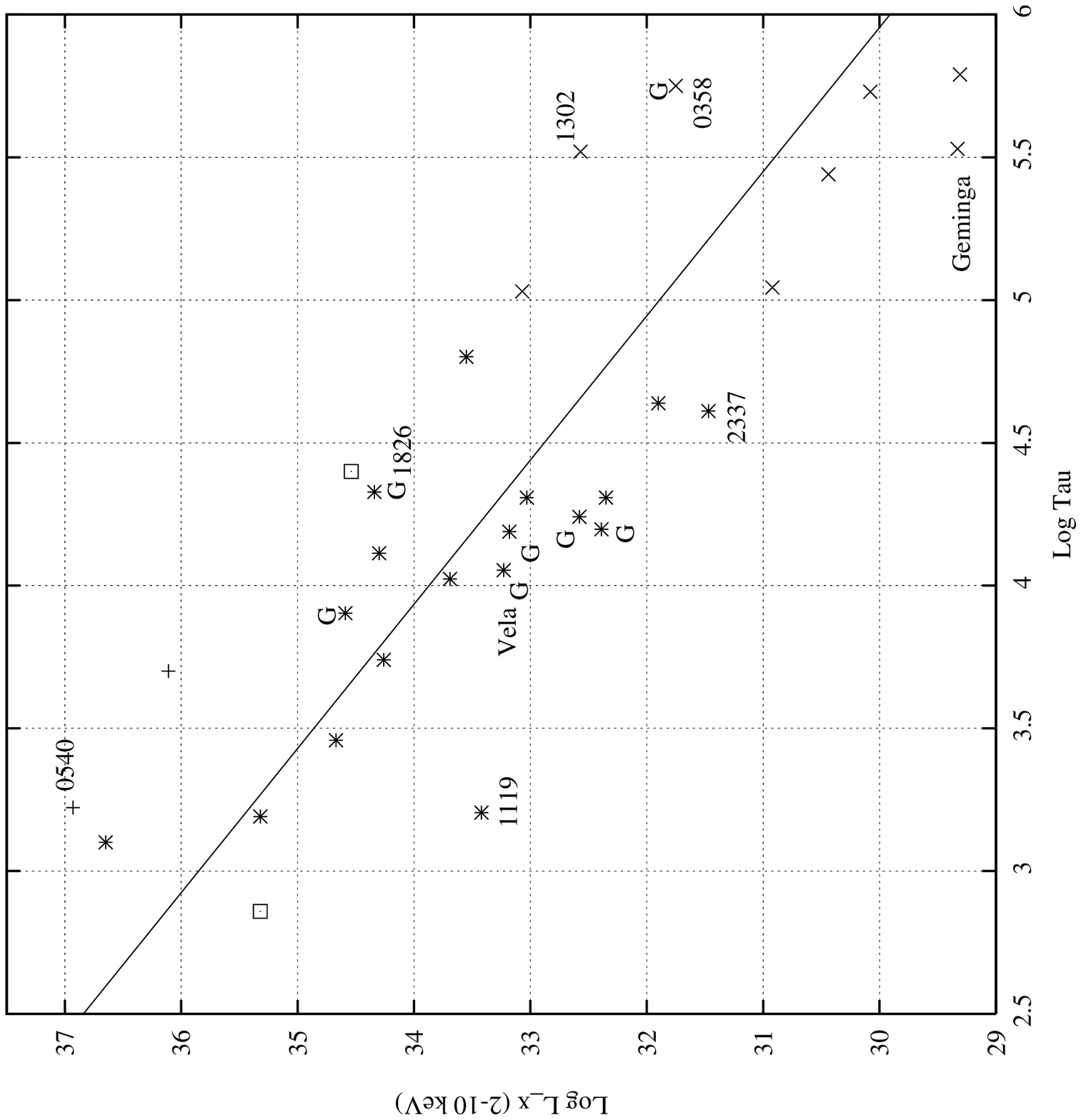}
\end{figure}

\clearpage
\begin{figure}[t]
\vspace{3cm}
\includegraphics{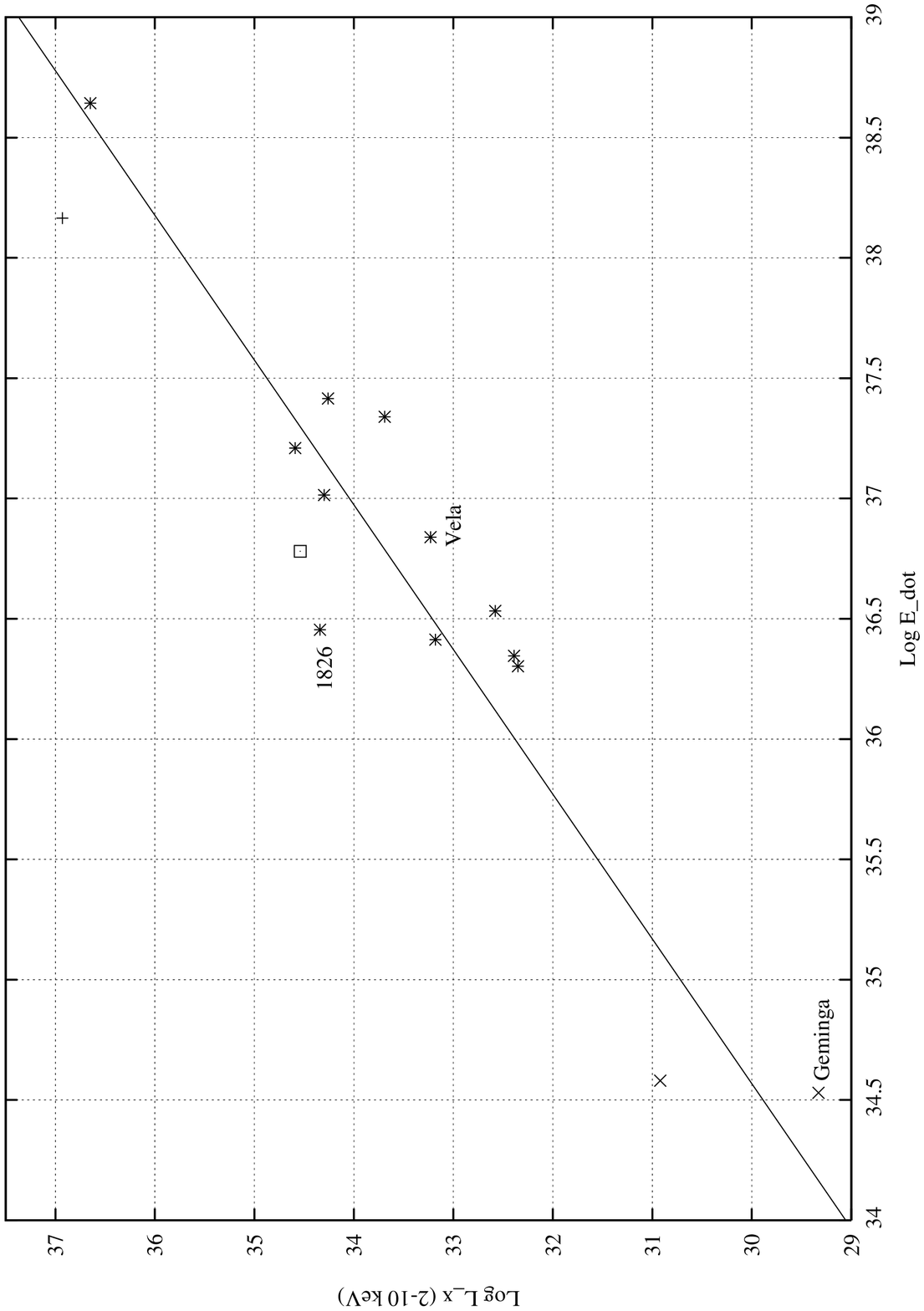}
\end{figure}

\clearpage
\begin{figure}[t]
\vspace{3cm}
\includegraphics{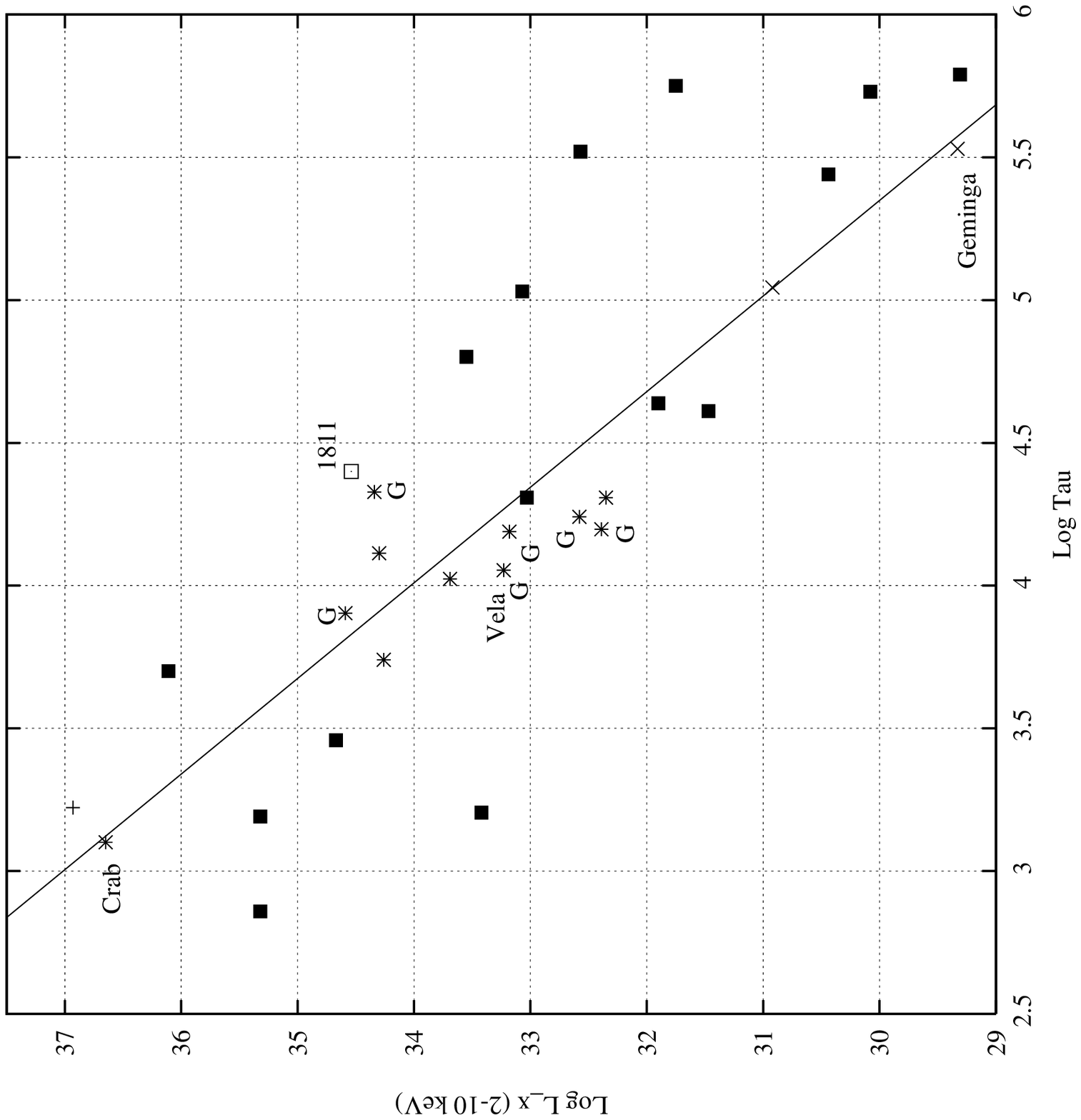}
\end{figure}

\clearpage
\begin{figure}[t]
\vspace{3cm}
\includegraphics{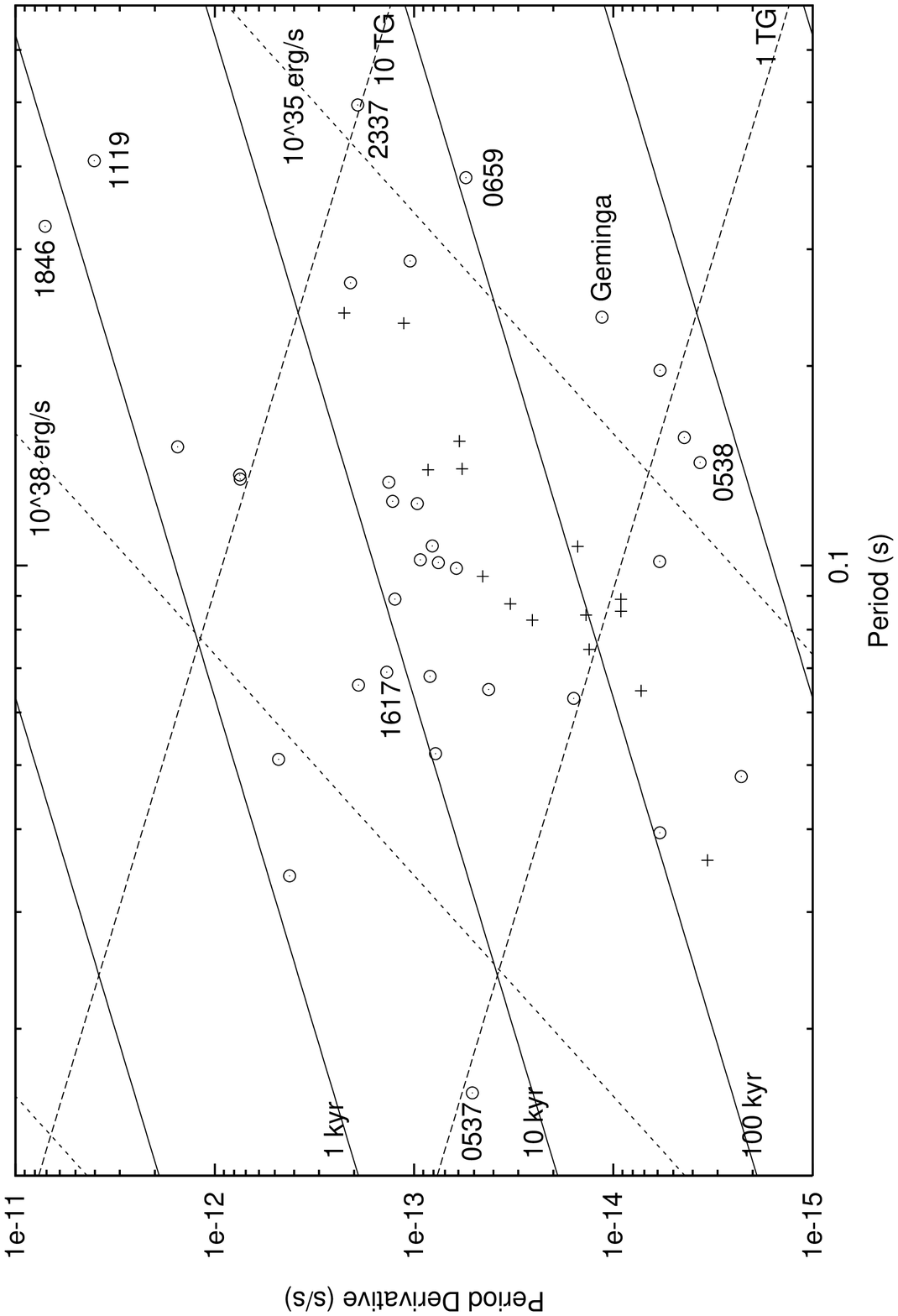}
\end{figure}

\clearpage
\begin{figure}[t]
\vspace{3cm}
\includegraphics{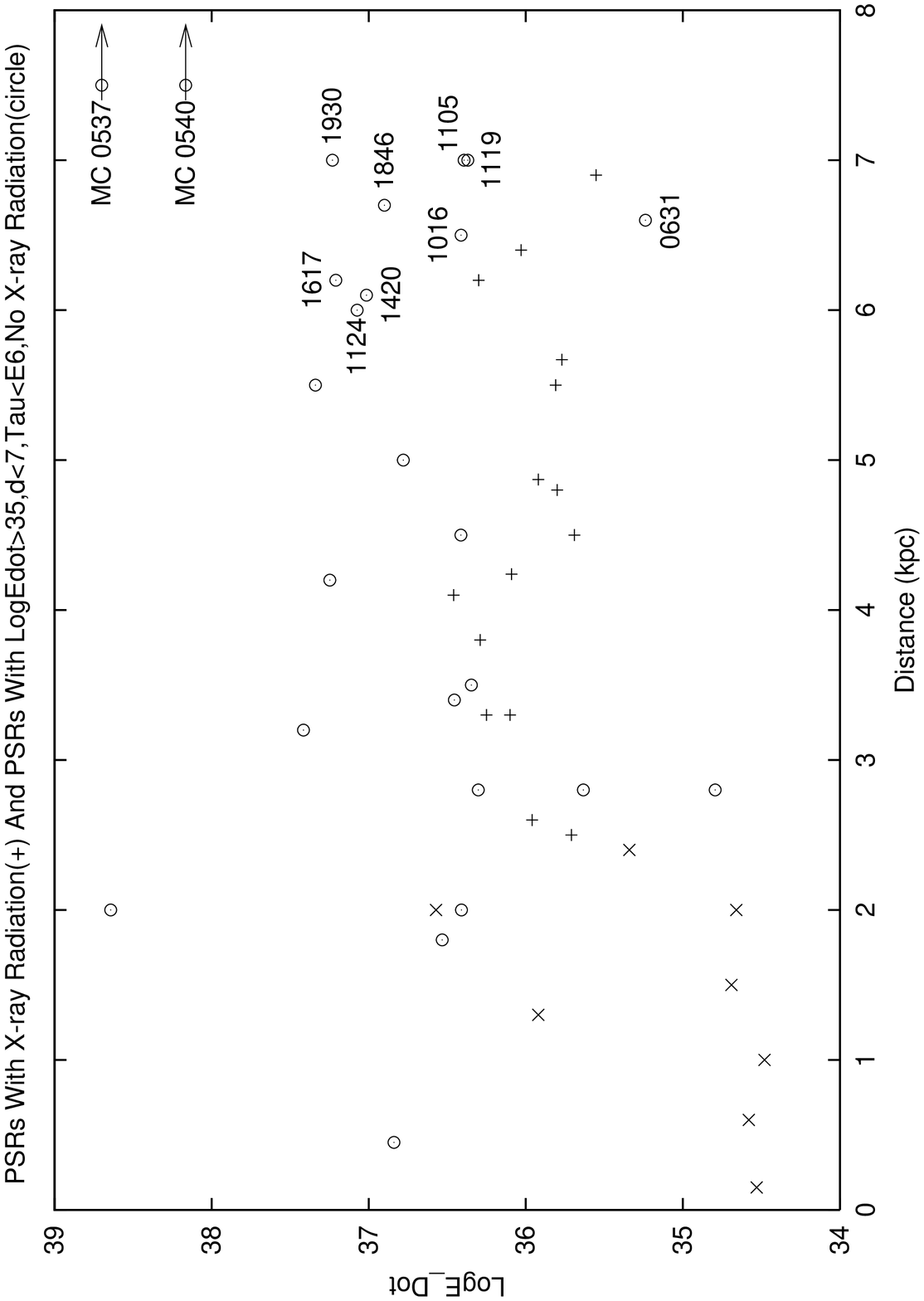}
\end{figure}

\clearpage  
\begin{figure}[t]
\vspace{3cm}
\includegraphics{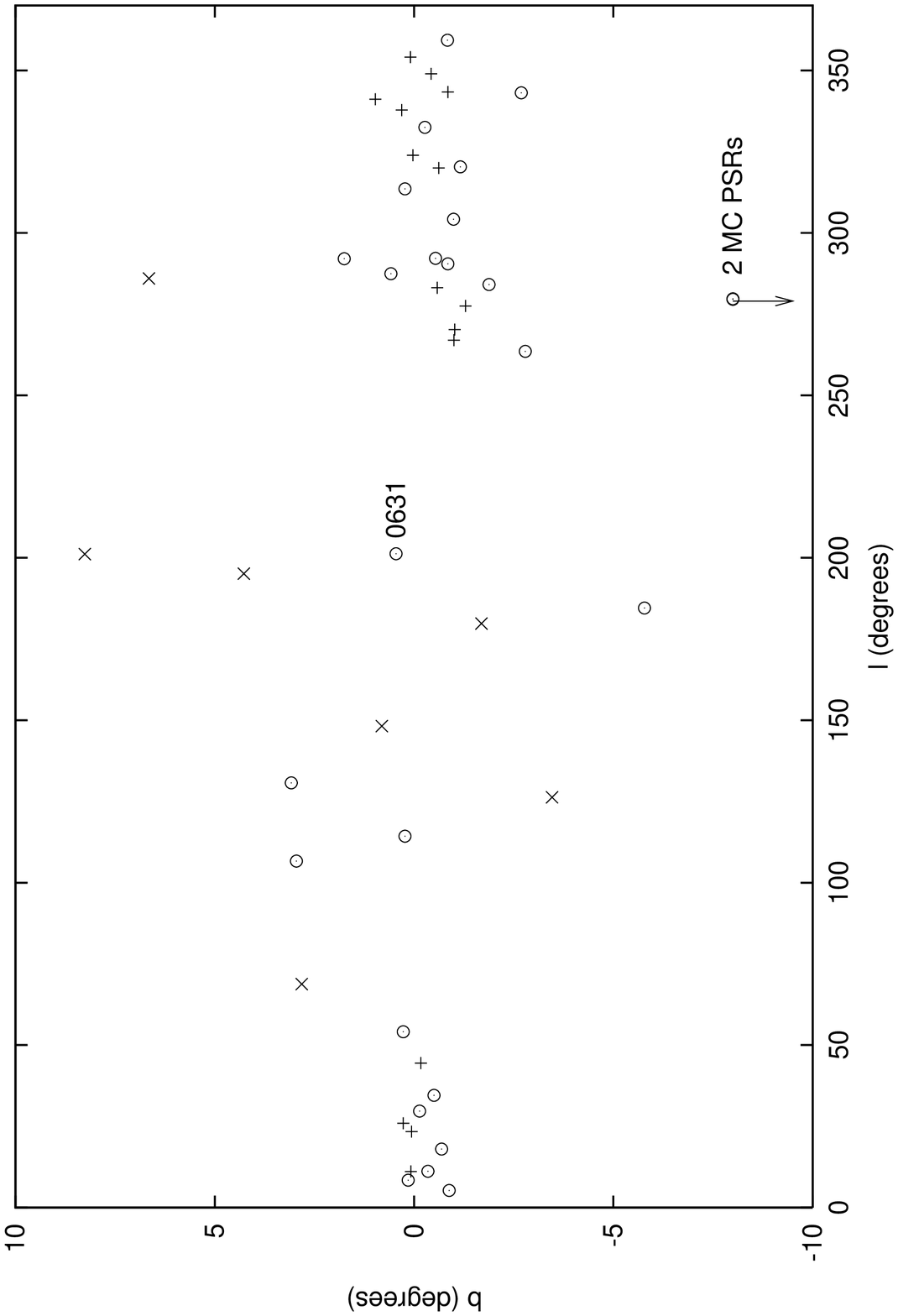}
\end{figure}

\end{document}